\documentclass[amssymb,amsmath,aps,twocolumn,prb,floatfix,showpacs]{revtex4}
\usepackage{graphicx}

\newcommand {\insfig}[2] {  \includegraphics[width=#2] {#1} }

\begin{document}
%\DeclareGraphicsExtensions{.eps}

\title{Interaction of vortices in superconductors with $\kappa $ close to $1/\sqrt{2%
}$}
\author{F.\ Mohamed} \altaffiliation[Actual address:]{CSCS, Galleria 2, CH-6928 Manno}
\email{fawzi@cscs.ch}
\author{M.\ Troyer}\author{G.\ Blatter}
\affiliation{Theoretische Physik, ETH-H\"onggerberg, CH-8093 Z\"urich,
Switzerland}
\author{I.\ Luk'yanchuk}
\affiliation{L.\ D.\ Landau Institute for Theoretical Physics, Moscow, Russia;}
\affiliation{Institut Laue Langevin, BP156 38042 Grenoble Cedex 9, France}

\begin{abstract}
Using a perturbative approach to the infinitely degenerate Bogomolnyi
vortex state for a superconductor with $\kappa =1/\sqrt{2}$, $T\rightarrow
T_{c}$, we calculate the interaction of vortices in a superconductor with
$\kappa $ close to $1/\sqrt{2}$.
We find, numerically and analytically, that depending on the material the interaction
potential between the vortices varies with decreasing $\kappa $ from purely
repulsive (as in a type-II superconductor) to purely attractive (as in a
type-I superconductor) in two different ways: either vortices form a bound
state and the distance between them changes gradually from infinity to zero, 
or this transition occurs in a discontinuous way as a result
of a competition between minima at infinity and zero. We study
the discontinuous transition between the vortex and Meissner states caused by the
non-monotonous vortex interaction and calculate the corresponding
magnetization jump.
\end{abstract}

\pacs {74.60.Ec, 74.20.De, 74.55.+h, 74.60.-w}

\date{\today}

\maketitle

\section{Introduction}

It is widely known that superconducting vortices repel each other in
superconductors of type-II and attract each other in superconductors of
type-I. The physical origin of this phenomenon is the competition between the magnetic
repulsion of the vortices (dominating in type-II
superconductors) and the gain in condensation energy of overlapping vortex cores producing an attractive interaction (dominating in type-I superconductors).
Within the Ginzburg-Landau (GL) approximation it can easily be seen that
the long-range asymptotic behavior of the vortex interaction changes its
sign at  $\kappa=1/\sqrt{2}$. The vortex interaction is \cite{Kr} 
\begin{equation}
U_{\text{int}}(l)=2\pi c^{2}(\kappa )K_{0}(l)-\frac{\pi }{\kappa ^{2}}d(\kappa
)^{2}K_{0}(\sqrt{2}\kappa l)\;, \label{assympt}
\end{equation}
where $K_{0}(l)$ is the modified Bessel function of zero order, and $c(\kappa)$ and $d(\kappa)$
are slowly varying functions of $\kappa$ that are equal at $\kappa=1/\sqrt{2}$.
The detailed profile of $U_{\text{int}}(l)$ at any $l$ when $\kappa $ goes through $1/\sqrt{2}$ 
was, however, calculated only in the 70's by 
Jacobs and Rebbi \cite{Jac} using a special symmetry of the $z$-invariant GL
equations at $\kappa =1/ \sqrt{2}$ discovered by Bogomolnyi \cite{Bog}.
According to Bogomolnyi, Jacobs and Rebbi (BJR) the GL energy at 
$\kappa =1/\sqrt{2}$ is degenerate with respect to any configuration of vortices.
The sign change of $U_{\text{int}}(l)$ at $\kappa =1/\sqrt{2}$ is an \textit{exact}
result of the GL theory: at $\kappa >1/\sqrt{2}$ the interaction is
purely repulsive, at $\kappa =1/\sqrt{2}$ vortices do not interact, and
at $\kappa <1/\sqrt{2}$ the interaction is purely attractive.

Experiments \cite{Hubner,krae}, however, reveal a more complex situation.
The interaction potential $U_{\text{int}}(l)$ close to $\kappa = 1/\sqrt{2}$
was sometimes found to be attractive at large distances and repulsive at
short ones. This non-monotonous profile of $U_{\text{int}}(l)$ manifests itself as a
discontinuous transition between vortex and Meissner states and by the
existence of intermediate mixed-state domains of bound vortices. 
The presence of a local minimum in $U_{\text{int}}(l)$ at $\kappa \sim 1/\sqrt{2}$ 
can be explained (as we do here) by taking into
account low-temperature corrections to the GL theory that modify the
almost flat profile of the interaction \cite{Jac3}. Another possibility is
to take into account the fluctuations and anisotropies in the vortex lattice,
that produce an attractive interaction of the Van der Waals type \cite{VdW}.

Several calculations based on an extended GL functional \cite{TewHc1}
were done to clarify this issue.  Jackobs \cite{Jac3} calculated
low temperature corrections to the long-range vortex interaction 
(\ref{assympt}) and found that vortices attract each other already in type-II superconductors. 
Based on his results Hubert \cite{Hubert} performed numerical calculations for a periodic
Abrikosov lattice of vortices and demonstrated the
non-monotonous behavior of the vortex interaction in a vortex lattice. These
calculations are consistent with numerical variational calculations by Brandt \cite
{BrandtLatAll} based on Gorkov equations and solved for vortex lattice configurarions at all possible values of 
$H$, $T$, and $\kappa $. Although these results reproduce a nontrivial behavior of 
$U_{\text{int}}(l)$ at $\kappa \sim 1/\sqrt{2}$, it is difficult to survey and
interpret them in a systematic way because of the cumbersome mathematics and
large number of terms in the extended GL functional.

A new method of calculating the properties of superconductors near $%
\kappa \sim 1/\sqrt{2}$ has been developed recently \cite{Lukyanc}. In this
approach the degenerate vortex state at $\kappa =1/\sqrt{2}$ is considered
as the starting point of a secular perturbation theory. A degeneracy-lifting 
perturbation functional in the small parameters
\begin{equation}
\gamma =\kappa ^{2}-\frac{1}{2},\qquad t=\frac{T}{T_{c}}-1  \label{small}
\end{equation}
is constructed. This approach avoids bulky calculations and allows to
describe the superconductor with $\left| \gamma \right| $,$\left| t\right|
\ll 1$ in a form that is easy to interpret.

We will use this perturbation approach to calculate, the interaction $U_{\text{int}}(l)$ between two separate vortices and
between vortices in a lattice, for a superconductor with $\kappa \sim 1/%
\sqrt{2}$ when the Ginzburg-Landau theory is extended to low temperatures.
We study the discontinuous transition between the vortex and Meissner states
caused by the nonmonotonous vortex interaction and calculate the
corresponding magnetization jump.

\section{Perturbation approach}

\label{perturb}

We start by outlining the main elements of our perturbative approach. \cite{Lukyanc}
According to Bogomolnyi \cite{Bog}, and Jacobs and Rebbi \cite{Jac}, in a $z$-invariant situation, 
the order parameter amplitude $\left| \psi (\mathbf{r},\mathbf{r}_{1},\ldots,%
\mathbf{r}_{n})\right| $ of vortices located at 
$\mathbf{r}=\mathbf{r}_{1},...,\mathbf{r}_{n}$ in a superconductor with $\gamma =0$, $%
t\rightarrow 0$ is described by the BJR equation
\begin{equation}
\frac{1}{2}\nabla ^{2}\ln \left| \psi \right| ^{2}=\left| \psi \right|
^{2}-1+\sum_{i}2\pi \delta \left( \mathbf{r}-\mathbf{r}_{i}\right).  \label{BJR}
\end{equation}
The magnetic field inside the sample is uniquely
related to $\left| \psi (\mathbf{r})\right| $ via
\begin{equation}
b(\mathbf{r})=1-\left| \psi (\mathbf{r})\right| ^{2}.  \label{field}
\end{equation}
Here, dimensionless variables $\psi =\Psi /\Psi
_{0}$, $b=\sqrt{2}\kappa B/H_{c}$ are used (with $\Psi _{0}$ the uniform order parameter when the external field $H=0$).
Distances are measured in units of the coherence length $\xi $.

Since at $\gamma =0$, $t\rightarrow 0$ the vortices do not interact the
vortex energy close to this point can be calculated to first order
in $\gamma $ and $t$ by substituting the unperturbed solution 
$\left| \psi (\mathbf{r},\mathbf{r}_{1},...,\mathbf{r}_{n})\right| $ of Eq.\ (\ref{BJR}) into the functional
\begin{equation}
f=(h_{0}-h_{c2})\left| \psi \right| ^{2}+(\gamma -c_{4}t)\left| \psi \right|
^{4}-c_{6}t\left| \psi \right|^{6} \label{Secul}
\end{equation}
as obtained in Ref.\ \onlinecite{Lukyanc} and defined for 
the class of Bogomolnyi solutions. Here $f=8\pi \kappa ^{2}{\cal F}/H_{c}^{2}+\kappa ^{2}$
(${\cal F}$ is the GL free energy) and $h_{c2}=\sqrt{2}\kappa H_{c2}/H_{c}$ are the dimensionless free energy and
the upper critical field. The functional (\ref{Secul}) accounts for all the terms
of the extended GL functional that are assembled into the terms $\left| \psi
\right| ^{4}$ and $\left| \psi \right| ^{6}$ with experimentally measurable
material coefficients $c_{4}$ and $c_{6}$. The parameter $c_{4}$ is always
positive, whereas $c_{6}$ can be both positive and negative.

To calculate the energy of vortices located at 
$\mathbf{r}=\mathbf{r}_{1},...,\mathbf{r}_{n}$ one should solve first the BJR
equation (\ref{BJR}) and then substitute this solution into the perturbation
functional (\ref{Secul}). The analytical aspects of this task have been
discussed in detail in Ref.\onlinecite{Lukyanc}; here we solve the BJR equation (\ref
{BJR}) numerically by a finite elements method on an adaptive grid by using the Ansatz
\begin{equation}
|\psi (\mathbf{r})|=\prod_{k}|\mathbf{r}-\mathbf{r}_{k}|e^{\varphi (\mathbf{r})/2}.
\label{An}
\end{equation}
Using this Ansatz, the BJR equations are reduced to a nonlinear Poisson-like equation for 
$\varphi $ and the delta-functions are removed. In a weak
formulation for the finite dimensional space ($\varphi (\mathbf{r}%
)=\sum_{i}\varphi _{i}p_{i}(\mathbf{r})$), the problem is written as
\begin{gather}
-\frac{1}{2}\sum_{j}\varphi _{j}\int_{\Omega }\nabla p_{j}\cdot \nabla p_{i}d%
\mathbf{r}=-\frac{1}{2}\int_{\partial \Omega }p_{i}\nabla \varphi \times d\mathbf{l%
}+  \nonumber \\*
\int_{\Omega }\left( e^{\sum_{j}p_{j}\varphi _{j}}\prod_{k}|\mathbf{r}-\mathbf{r}%
_{k}|^{2}-1\right) p_{i}d\mathbf{r},  \label{femPde}
\end{gather}
where, $p_{j}(\mathbf{r})$ are Lagrange elements on a quadratic grid \cite
{deal}. We linearize the exponential function and iterate the
resulting linear part. The dynamic refinement is based on the normal Kelly
indicator \cite{ains} for the local error and on the error evaluation
of the nonlinear exponential part.

As boundary conditions we use either Neumann (specifying $\partial _{\perp }|\psi |$) or
periodic boundary conditions. In the Neuman case the normal derivative of $%
\varphi $ is taken to be $\partial _{\perp }\varphi =-2\sum_k \partial _{\perp
}\ln (|\mathbf{r}-\mathbf{r}_k|)$. In the periodic case a special constraint is
added to the system of linear equations for $\varphi_i \in \mathbb{R}$, to
compensate for the nonperiodicity of $\sum_k\ln (|\mathbf{r}-\mathbf{r}_{k}|)$.
This algorithm is implemented in C++ using the deal-II library \cite{deal} 
for the finite elements calculations.

\section{Two vortex interaction}

We now use the above method to calculate the interaction energy $U_{\text{int}}(l)$
between two vortices located at $\mathbf{r}_{1,2}=\pm \mathbf{l}/2$ by
subtracting the self-energy of separated vortices $2\varepsilon _{1}$ 
from the two-vortex energy $\varepsilon _{1,1}(l)$,
\begin{equation}
U_{\text{int}}(l)=\varepsilon _{1,1}(l)-2\varepsilon _{1}=\varepsilon
_{1,1}(l)-\varepsilon _{1,1}(\infty ).  \label{twi}
\end{equation}
As follows from (\ref{Secul}) the energy $U_{\text{int}}(l)$ can be written as a
superposition of two structure functions $u_{k}(l)$, $k=4,6$
\begin{equation}
U_{\text{int}}(l)=(\gamma -c_{4}t)\,u_{4}(l)-c_{6}t\,u_{6}(l),  \label{superp}
\end{equation}
which do not depend on the material parameters $\gamma $, $c_{4}$ and $c_{6}$,
\begin{eqnarray}
u_{k}(l) &=&\int [(1-\left| \psi (\mathbf{r},\mathbf{r}_{1},\mathbf{r}_{2})\right| ^{k})-(1-\left| \psi (\mathbf{r},\mathbf{r}_{1})\right| ^{k})
\nonumber \\*
&&-(1-\left| \psi(\mathbf{r},\mathbf{r}_{2})\right| ^{k})]dS.
\end{eqnarray}
Here $\psi (\mathbf{r},\mathbf{r}_{1,2})$ and $\psi (\mathbf{r},\mathbf{r}_{1},\mathbf{r}%
_{2})$ are the one- and two-vortex solutions of (\ref{BJR}). Note that \cite{Lukyanc} 
$u_{2}(l)=0$.

It follows from (\ref{superp}) that the profile of $U_{\text{int}}(l)$ depends only on the
sign of $c_{6}$ and on the control parameter 
\begin{equation}
d=\frac{\gamma -c_{4}t}{\left| c_{6}t\right| }.  \label{d}
\end{equation}

\begin{figure}[htbp]
\insfig{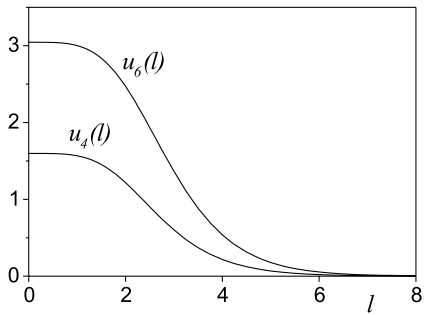}{5.5cm}
\vspace{0.0cm}
\caption{
Structure functions $u_{4}(l)$ and $u_{6}(l)$ for two vortices separated by
a distance $l$. Appropriate superposition of $u_{4}(l)$ and $u_{6}(l)$ gives the vortex
interaction energy $U_{\text{int}}(l)$. The distance $l$ is measured in units of  the
coherence length $\xi$.}
\label{TwovFigu12}
\end{figure}

In Fig.\ \ref{TwovFigu12} we show the result of our numerical calculations of
$u_{4}(l)$ and $u_{6}(l)$.

With the numerical method outlined in Sec.\ref{perturb} we can calculate
the minimum of $U_{\text{int}}$ for $1.4\leq d\leq 2.7$. For $d<1.4$ and for $%
d>2.7$ the minimum cannot be reliably found because of the flat profile
of $U_{\text{int}}(l)$ at small and large values of $l$. To determine the behavior
of $U_{\text{int}}(l)$ at $l\ll 1$ and at $l\gg 1$ we have used the analytic
estimates of Ref.\ \onlinecite{Lukyanc} that are summarized below:

\textit{i)} the order parameter of the widely separated weakly overlapping vortices 
($l\gg 1$) can be approximated as
\begin{equation}
\left| \psi (\mathbf{r})\right| ^{2}=g_{1}^{2}(\mathbf{r}+\mathbf{l}/2)+g_{1}^{2}(%
\mathbf{r}-\mathbf{l}/2)-1,  \label{wide}
\end{equation}
where $g_{1}(\mathbf{r})$ is the axially symmetric one-quantum vortex solution
of the BJR equation. The long-range interaction is written as (up to the slow
pre-exponential factor $u(l)$)
\begin{equation}
U_{\text{int}}(l)=[\gamma -(c_{4}+3c_{6})t]\cdot u(l)e^{-4l},  \label{Ularge}
\end{equation}
and becomes attractive when $\gamma $ is smaller than the critical value 
\begin{equation}
\gamma _{c1}=(c_{4}+d_{c1}c_{6})t,\qquad d_{c1}=3.  \label{gc1}
\end{equation}
Eq.\ (\ref{Ularge}) has been obtained before in Ref.\ \onlinecite{Jac3} and generalizes Eq.
(\ref{assympt}) to lower temperatures.

\textit{ii)} the order parameter of two close-lying vortices with almost coinciding
cores ($l\ll 1$) can be approximated as
\begin{equation}
\left| \psi (\mathbf{r})\right| =g_{2}(r)+\frac{1}{8}g_{2}(r)(\mathbf{l}\nabla %
)^{2}\ln g_{2}(r),  \label{close}
\end{equation}
where $g_{2}(\mathbf{r})$ is the axially symmetric two-quantum vortex solution
of the BJR equation. At small $l$ the short-range vortex interaction is
expanded as
\begin{equation}
U_{\text{int}}(l)=[0.91(\gamma -c_{4}t)-1.13c_{6}t]\,l^{2}+O(l^{4}).
\label{Usmall}
\end{equation}
It is attractive when the second order term is negative, i.e.,
when $\gamma $ is smaller than the critical value 
\begin{equation}
{\gamma }_{c2}=(c_{4}+d_{c2}c_{6})t,\qquad d_{c2}=1.26.  \label{gc2}
\end{equation}

We finally compare the energy of a two-quanta vortex to the energy
of two widely separated one-quantum vortices. A two-quanta vortex 
becomes energetically more favorable when $\gamma $ is
smaller than the critical value
\begin{equation}
\gamma _{c}=(c_{4}+d_{c}c_{6})t,\qquad d_{c}=1.89.  \label{gc}
\end{equation}
This value always lies in between $\gamma _{c1}$ and $\gamma _{c2}$.
The critical parameter $d_{c1}$ remains unchanged in the
case of interacting multi-quanta vortices whereas the numerical coefficients 
$d_{c2}$ in Eq.\ (\ref{gc2}) and $d_{c}$ in Eq.\ (\ref{gc}) become larger.

Combining these numerical and analytical results we conclude that
depending on the parameters $c_{6}\,\ $and $c_{4}$ two scenara are possible
for the evolution of $U_{\text{int}}(l)$ as function of $d$:

\begin{figure}[tbp]
\insfig{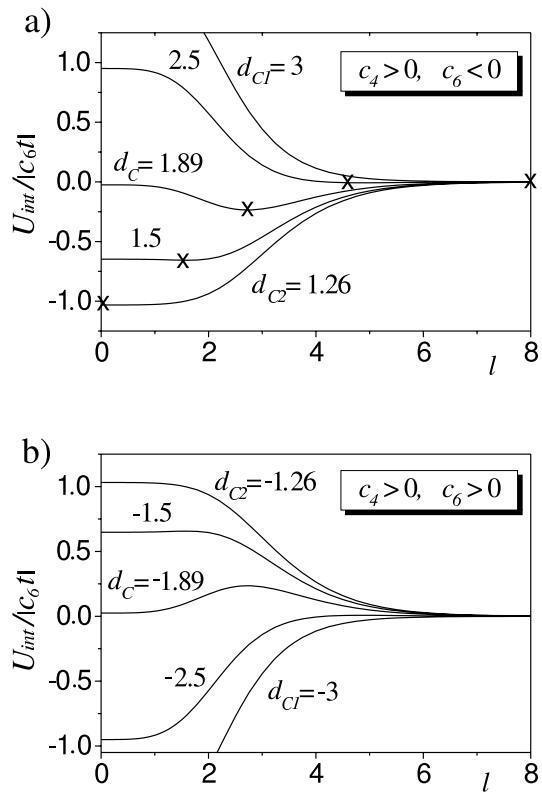}{7cm}
\vspace{0.0cm}
\caption{The reduced two-vortex interaction energy $U_{\text{int}}(l)/\left|
c_{6}t\right| $ at different parameters $d=(\gamma
-c_{4}t)/\left| c_{6}t\right| $ (a) for $c_{4}>0$, \ $c_{6}<0$ (crosses
indicate the minima) and (b) for $c_{4}>0$, \ $c_{6}>0$.}
\label{TwovFigUint}
\end{figure}

I. The case $c_{4}>0$, \ $c_{6}<0$ is shown in Fig.\ \ref{TwovFigUint}a. When 
$d>d_{c1}=3$ the interaction is purely repulsive. Below $d_{c1}$ we find an 
attraction at large distances while the short-range interaction remains repulsive. 
The vortices form a bound state with an equilibrium distance $l_{0}(d)$ corresponding
to the minimum of $U_{\text{int}}(l)$. Finally, below the second critical value $d_{c2}=1.26$, the
short-range interaction becomes attractive too and the vortices combine into
a two-quanta vortex. 

\begin{figure}[htbp]
\insfig{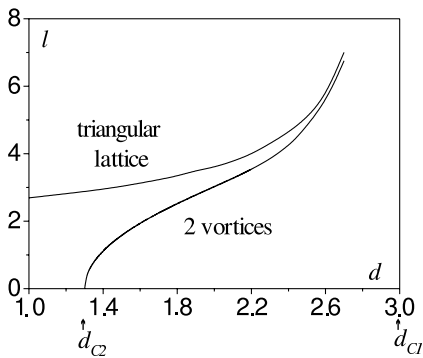}{5.5cm}
\vspace{0.0cm}
\caption{
Distance between two bound vortices as a function of the material parameter $d$ (when $%
c_{4}>0$, \ $c_{6}<0$). The equilibrium vortex distance in the regular
triangular vortex lattice at discontinuous transition at $H=H_{c1}^{\ast }$
is shown for comparison.}
\label{TwovFigL}
\end{figure}

The distance $l_{0}(d)$ of two bound vortices, shown in Fig.\ \ref {TwovFigL}, 
diverges as $-\ln (d_{c1}-d)$
for $d\rightarrow d_{c1}$ and vanishes as $\sqrt{d-d_{c2}}$ for $d\rightarrow
d_{c2}$. Special care has to be taken to observe these vortex bound states.
One can try, e.g., to obtain pinned vortex molecules after expelling other
vortices from the sample by turning off the external field.

II. The case $c_{6}>0$, $c_{4}>0$ is shown in Fig.\ \ref{TwovFigUint}b. The
order of the critical parameters $d_{c2}$, $d_{c1}$ is now reversed, 
$d_{c1}<d_{c2}$. 
The vortex interaction now changes its sign at negative $d$.
When $d>d_{c2}=-1.26$ the interaction is purely repulsive. Below $d_{c2}$
the interaction becomes short-range attractive while retaining its long-range
repulsive character. Below $d_{c1}=-3$ the interaction is purely attractive.
The equilibrium vortex configuration in the interval $d_{c1}<d<d_{c2}$ 
is thus determined by the competition between
two local minima of $U_{\text{int}}(l)$ at $l=0$ and at $l=\infty $. The formation of
a two-quanta vortex becomes more favorable than two isolated and widely
separated vortices below the critical value $d_{c}=-1.89$. The parameters $%
d_{c1}$ and $d_{c2}$ then serve as supercooling and superheating
limits of $d$.

The cases considered above exhaust all possible scenaria for the evolution of 
$U_{\text{int}}(l)$ near the Bogomolnyi point. Case I seems to be more realistic given
the typical values of the material constants \cite{Lukyanc}: $c_{4}\sim 0.1-0.5$, 
$c_{6}\sim -0.2-0$.

We conclude this section by remarking that 
the vortex-antivortex interaction cannot be
calculated in this formalism since the
vortex-antivortex pair does not belong to the degenerate set of the
Bogomolnyi states at $\kappa =1/\sqrt{2}$. Moreover, unlike the case of
parallel vortices, both the magnetic and the condensation energy contribute equally to the
attractive vortex-antivortex interaction and thus there is
no critical point at $\kappa =1/\sqrt{2}$.

\section{Vortex lattices}

The energy of an ensemble of vortices is a nonlinear function of the vortex
positions. It can be reduced to the pairwise vortex interaction (\ref{twi}) only
in the case of large vortex separation. Then, in the nearest-neighbor
approximation, the energy of the regular lattice of one-quantum vortices with
lattice constant $l\gg 1$, calculated with respect to the uniform Meissner
state, can be written as: 
\begin{equation}
f_{\text{lat}}=\frac{\overline{b}}{2\pi }\varepsilon _{1}+\frac{\overline{b}}{2\pi }%
\frac{z}{2}U_{\text{int}}(l),  \label{dillatenfirst}
\end{equation}
where $\varepsilon _{1}$ \ is the one-vortex energy, $z$ is the coordination
number, $\overline{b}/2\pi $ is the vortex density. The average magnetization
$\overline{b}$ of the lattice is related to the unit
cell area $S_{l}$ of the vortex lattice: 
\begin{equation}
\overline{b}=2\pi /S_{l}.
\end{equation}
For equilateral triangular (with $z=6$) and square (with $z=4$) lattices it
is: $\overline{b}_{\square }=2\pi /l^{2}$, $\overline{b}%
_{\Delta }=4\pi /l^{2}\sqrt{3}$.

At closer distance between vortices the energy $f_{\text{lat}}$ is renormalized
both because of the next-nearest neighbor vortex interaction and because of
the nonlinear corrections. The first contribution can be accounted for by
a redefinition of $U_{\text{int}}(l)$ as $U_{\text{int}}(l)+U_{\text{int}}(\sqrt{3}l)$ for a triangular and $U_{\text{int}}(l)+U_{\text{int}}(\sqrt{2}l) $ for a square lattice.

However, the nonlinear corrections can only be calculated by a
vortex-lattice solution of the BJR equation (\ref{BJR}).
In fact one extend Eq.\ (\ref{dillatenfirst}) to the case
of arbitrary $l$ provided that the contribution $(z/2) U_{\text{int}}(l)$ 
is substituted with the interaction energy $(z/2) U_{\text{lat}}(l)$ calculated from the numerical solution of the BJR equation for a periodic
vortex lattice. Similar to the case of two vortices, the energy 
 can be expressed as
\begin{equation}\label{eqUlat}
\frac{z}{2}U_{\text{lat}}(l)=(\gamma -c_{4}t)\,v_{4}(l)-c_{6}t\,v_{6}(l)
\end{equation}
via the lattice-dependent structure functions $v_{k}(l)$,
\begin{gather}
v_{k}(l)=\int \left( 1-\left| \psi _{l}(\mathbf{r})\right| ^{k}\right)
dS_{l}-\nonumber \\*
\int \left( 1-\left| \psi _{\infty }(\mathbf{r})\right| ^{k}\right)
dS_{\infty }.  \label{ukl}
\end{gather}

\begin{figure}[htbp]
\insfig{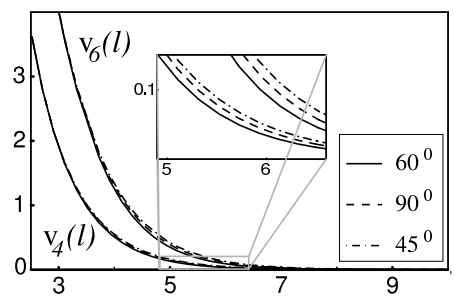}{6cm}
\vspace{0.0cm}
\caption{
Structure functions $v_{4}(l)$ and $v_{6}(l)$ for the square ($90^0$), triangular ($60^0$) 
and rhombic ($45^0$) lattices with lattice constant 
 $l$. Superposition of $v_{4}(l)$ and $v_{6}(l)$ gives the lattice interaction function  
 $(z/2) U_{\text{lat}}(l)$. The distance is measured in units of  the
coherence length $\xi$.}
\label{TwovFigv12}
\end{figure}

The vortex lattice order parameter $\left| \psi _{l}(\mathbf{r})\right| $ is
obtained from a numerical solution of Eq.\ (\ref{BJR}),
applying the method discussed in Sec. \ref{perturb} to a unit cell $S_{l}$ 
with periodic boundary conditions. 
The one-vortex solution $\left| \psi _{\infty }(\mathbf{r})\right| $ is calculated
on an infinite unit cell $S_{\infty }$. In Fig.\ \ref{TwovFigv12} we show as an example
the structure functions $v_{k}(l)$ for the triangular, square, and rhombic lattices.

\begin{figure}[htbp]
\insfig{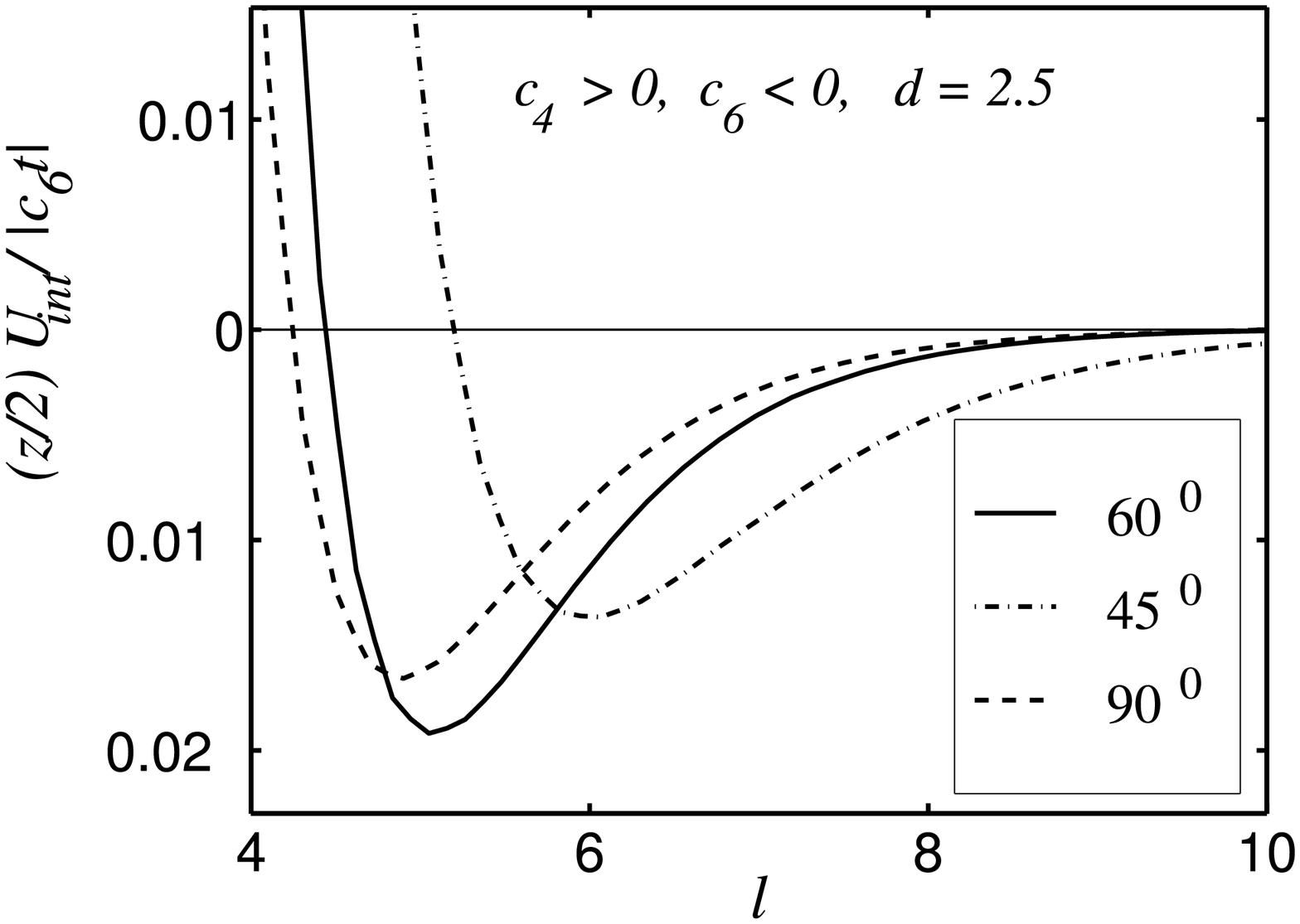}{6cm}
\vspace{0.0cm}
\caption{
The reduced vor\-tex in\-te\-rac\-tion ener\-gy 
$(z U_{\text{lat}}(l))/\\* (2 \left| c_6 t \right|)$ in square ($90^0$), trian\-gu\-lar ($60^0$), 
and rhom\-bic ($45^0$) lattices at $d=2.5$.}
\label{TwovFigUlat}
\end{figure}

\begin{figure}[htbp]
\insfig{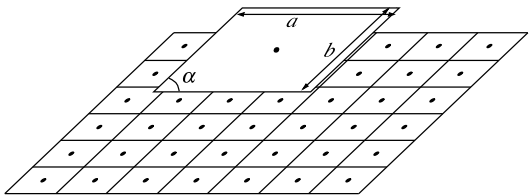}{7cm}
\vspace{0.0cm}
\caption{
Possible vortex lattice structures considered here, $1/6<a/b\leq 1$, $\pi/12\le \alpha \le \pi/2$.
}
\label{latticeFig}
\end{figure}

The vortex interaction $(z/2) U_{\text{lat}}(l)$ depends on the
control parameter $d$. Here we limit our discussion to the most realistic case $c_{4}>0$, $c_{6}<0$. We find a nonmonotonous behavior at $d<d_{c1}=3$ with
long-range attraction, short-range repulsion and a minimum at intermediate
distances as
shown in Fig.\ \ref{TwovFigUlat} (thus we deal with the usual type II superconductor case when $d>3$, and the type I superconductor when $d<1$). We have considered all the simple lattices of the form shown in Fig.\ \ref{latticeFig}, and, in a non-systematic way, some other lattices (multi-quanta, and 4-cluster lattices). Among these, the lowest
energy has always been found for a triangular lattice.

The most important consequence of the nonmonotonous vortex interaction is
the existence of a special class of superconductors that are intermediate
between type-I and type-II (we call them type I/II). They are
characterized by a discontinuity of the transition between the Meissner and
vortex states. A survey of properties of these superconductors has been given
in our previous publication \cite{Lukyanc}, where we have demonstrated their
existence in the interval $1<d<3$. We discuss now the details of the
Meissner-vortex transition in type-I/II superconductors based on the
the expression (\ref{eqUlat}) calculated above.

\begin{figure}[htbp]
\insfig{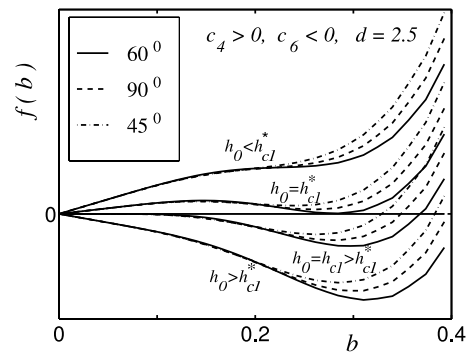}{6cm}
\caption{Energy $f$ of the different vortex lattices as a function of 
$\overline b$ in an external field $h_0$. The minimum 
of $f(\overline b)$ corresponds to the most stable vortex state at a given $h_0$.
The Meissner state corresponds to $\overline b =0$. The lattice constant $l$ is 
inversely proportional to $\overline b^{1/2}$. The material parameter $d$ is the
same as in Fig.\ \ref{TwovFigUlat}.}
\label{TwovFigEn}
\end{figure}

The stability of the vortex lattice in an external magnetic field $h_{0}$ can be
investigated by minimizing
\begin{eqnarray}
f(\overline{b})&=&f_{\text{lat}}(\overline{b})-f_s= \nonumber \\*
&&\frac{\overline{b}}{%
2\pi }\left[ \frac{z}{2}U_{\text{lat}}(l(\overline{b}))-2\pi (h_{0}-h_{c1})\right] 
\label{fm}
\end{eqnarray}
($f_s$ is the Meissner free energy) over $\overline{b}$ and different lattice types.
As can be seen in Fig.\ \ref{TwovFigEn} the transition from
Meissner state with $\overline{b}=0$ to the vortex state with finite $\overline{b}$ 
in type-I/II superconductors occurs discontinuously at a
critical field $h_{c1}^{\ast }$ that is smaller than the field
$h_{c1}=\varepsilon _{1}/2\pi$
where the penetration an individual vortex becomes energetically
advantageous. The critical field $h_{c1}$ which serves as the lower critical field 
of the continuous vortex--Meissner state transition in type-II superconductors can be interpreted as the field of superheating for type-I/II superconductors.

Another important conclusion is that,
among the different types of lattices studied here, the triangular one-quantum lattice is
the most stable one. We can show this numerically for all values of $d$ and $h_{0}$
except for the interval $2.7<d<3$, $h_{0}\sim h_{c1}^{\ast }$, where due to
the large distance between vortices the accuracy of our calculation has been insufficient to draw definitive conclusions.

\begin{figure}[htbp]
\insfig{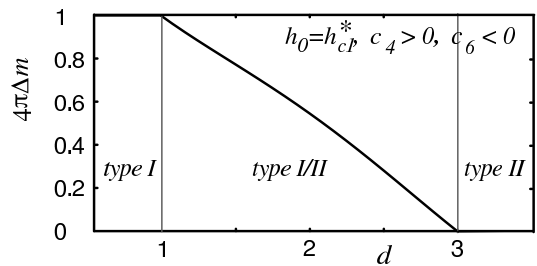}{7cm}
\vspace{0.0cm}
\caption{
Jump of magnetization $4\pi \Delta m=\overline{b}%
(h_{c1}^{\ast })$ at the transition between vortex and Meissner state as function of the control parameter $d$
 (case $c_{4}>0$, \ $c_{6}<0$).
}
\label{TwovFigJump}
\end{figure}

In Fig.\ \ref{TwovFigJump} we show the magnetization jump 
$4\pi \Delta m=\overline{b}(h_{c1}^{\ast })$ at the $h_{c1}^{\ast }$ transition as a function of 
$d$. Flux expulsion varies from almost complete as in type-I
superconductors at $d=1$ to vanishingly small at $d=3$ as in type-II
superconductors. Surprisingly, we find that the jump $\Delta m$ is almost linearly dependent on $d$ between
these two values. We also show the equilibrium lattice constant of the triangular lattice 
$l_{\Delta }=(\sqrt{3}\Delta z)^{-1/2}$ in Fig.\ Fig.\ \ref{TwovFigL} in order to compare it with the distance
between vortices bound in a pair. At $d=1$ the distance
between vortices in a lattice is minimal and equal to $l_{\Delta }=(4\pi /\sqrt{3})^{1/2}
\approx 2.69$. The vortex distance in the lattice is larger than for the vortex pair, because the presence of other vortices suppresses 
superconductivity and thus diminishes the pressure of the superconducting phase against 
the vortices. At $d\rightarrow d_{c1}=3$ the vortex distance
diverges and the lattice energy can be approximated as a superposition of pairwise interaction energies.

The discontinuity at the $h_{c1}^{\ast }$ transition can be observed experimentally
as a spinodal vortex clustering in the intermediate/mixed state
in thin superconducting plates \cite{Hubner}. The properties of such
clusters are determined by the delicate balance between the non-monotonous
vortex interaction $U_{\text{lat}}(l)$ and the vortex repulsion due to the additional interaction through the stray magnetic field energy in the vacuum outside the superconducting plate. Detailed calculations of this effect based on the
explicit form of $U_{\text{lat}}(l)$ are currently in a progress.

\begin{acknowledgments}
The authors are grateful to Dr.\ V.\ Geshkenbein for helpful discussions.
F.\ M.\ and I.\ L.\ thank Prof.\ H.\ Capellmann for his hospitality in RWTH-Aachen
where part of the work was done.
\end{acknowledgments}

\bibliography{Twov}

\begin{thebibliography}{13}
\expandafter\ifx\csname natexlab\endcsname\relax\def\natexlab#1{#1}\fi
\expandafter\ifx\csname bibnamefont\endcsname\relax
  \def\bibnamefont#1{#1}\fi
\expandafter\ifx\csname bibfnamefont\endcsname\relax
  \def\bibfnamefont#1{#1}\fi
\expandafter\ifx\csname citenamefont\endcsname\relax
  \def\citenamefont#1{#1}\fi
\expandafter\ifx\csname url\endcsname\relax
  \def\url#1{\texttt{#1}}\fi
\expandafter\ifx\csname urlprefix\endcsname\relax\def\urlprefix{URL }\fi
\providecommand{\bibinfo}[2]{#2}
\providecommand{\eprint}[2][]{\url{#2}}

\bibitem[{\citenamefont{Kramer}(1971)}]{Kr}
\bibinfo{author}{\bibfnamefont{L.}~\bibnamefont{Kramer}},
  \bibinfo{journal}{Phys. Rev.} \textbf{\bibinfo{volume}{B3}},
  \bibinfo{pages}{3821} (\bibinfo{year}{1971}).

\bibitem[{\citenamefont{Jacobs and Rebbi}(1979)}]{Jac}
\bibinfo{author}{\bibfnamefont{L.}~\bibnamefont{Jacobs}} \bibnamefont{and}
  \bibinfo{author}{\bibfnamefont{C.}~\bibnamefont{Rebbi}},
  \bibinfo{journal}{Phys. Rev.} \textbf{\bibinfo{volume}{B19}},
  \bibinfo{pages}{4486} (\bibinfo{year}{1979}).

\bibitem[{\citenamefont{Bogomolnyi}(1976)}]{Bog}
\bibinfo{author}{\bibfnamefont{E.~B.} \bibnamefont{Bogomolnyi}},
  \bibinfo{journal}{Yad. Fiz.} \textbf{\bibinfo{volume}{24}},
  \bibinfo{pages}{861} (\bibinfo{year}{1976}), \bibinfo{note}{see also [Sov. J.
  Nucl. Phys. \textbf{24}, 449 (1976)]; E. B. Bogomolnyi and A. I. Vainstein,
  Yad. Fiz. \textbf{23}, 1111 (1976) [Sov. J. Nucl. Phys. \textbf{23}, 588
  (1976)]}.

\bibitem[{\citenamefont{H\"{u}bener}(1979)}]{Hubner}
\bibinfo{author}{\bibfnamefont{R.~P.} \bibnamefont{H\"{u}bener}},
  \emph{\bibinfo{title}{Magnetic Flux Structures in Superconductors}}
  (\bibinfo{publisher}{Springer}, \bibinfo{address}{Berlin},
  \bibinfo{year}{1979}), \bibinfo{note}{for a review of experiments}.

\bibitem[{\citenamefont{Kr\"{a}geloh}(1970)}]{krae}
\bibinfo{author}{\bibfnamefont{U.}~\bibnamefont{Kr\"{a}geloh}},
  \bibinfo{journal}{Phys. Status. Solidi} \textbf{\bibinfo{volume}{42}},
  \bibinfo{pages}{559} (\bibinfo{year}{1970}).

\bibitem[{\citenamefont{Jacobs}(1971)}]{Jac3}
\bibinfo{author}{\bibfnamefont{A.~E.} \bibnamefont{Jacobs}},
  \bibinfo{journal}{Phys. Rev.} \textbf{\bibinfo{volume}{B4}},
  \bibinfo{pages}{3022} (\bibinfo{year}{1971}), \bibinfo{note}{see also A. E.
  Jacobs, Phys. Rev. \textbf{B4}, 3029 (1971)}.

\bibitem[{\citenamefont{Blatter and Geshkenbein}(1996)}]{VdW}
\bibinfo{author}{\bibfnamefont{G.}~\bibnamefont{Blatter}} \bibnamefont{and}
  \bibinfo{author}{\bibfnamefont{V.}~\bibnamefont{Geshkenbein}},
  \bibinfo{journal}{Phys. Rev. Lett.} \textbf{\bibinfo{volume}{77}},
  \bibinfo{pages}{4958} (\bibinfo{year}{1996}).

\bibitem[{\citenamefont{Neumann and Tewordt}(1966)}]{TewHc1}
\bibinfo{author}{\bibfnamefont{L.}~\bibnamefont{Neumann}} \bibnamefont{and}
  \bibinfo{author}{\bibfnamefont{L.}~\bibnamefont{Tewordt}},
  \bibinfo{journal}{Z. Physik} \textbf{\bibinfo{volume}{189}},
  \bibinfo{pages}{55} (\bibinfo{year}{1966}).

\bibitem[{\citenamefont{Hubert}(1972)}]{Hubert}
\bibinfo{author}{\bibfnamefont{A.}~\bibnamefont{Hubert}},
  \bibinfo{journal}{Phys. Stat. Sol. (b)} \textbf{\bibinfo{volume}{53}},
  \bibinfo{pages}{147} (\bibinfo{year}{1972}).

\bibitem[{\citenamefont{Brandt}(1976)}]{BrandtLatAll}
\bibinfo{author}{\bibfnamefont{E.~H.} \bibnamefont{Brandt}},
  \bibinfo{journal}{Phys. Stat. Sol. (b)} \textbf{\bibinfo{volume}{77}},
  \bibinfo{pages}{105} (\bibinfo{year}{1976}).

\bibitem[{\citenamefont{Luk'yanchuk}(2001)}]{Lukyanc}
\bibinfo{author}{\bibfnamefont{I.}~\bibnamefont{Luk'yanchuk}},
  \bibinfo{journal}{Phys. Rev.} \textbf{\bibinfo{volume}{B63}},
  \bibinfo{pages}{174504} (\bibinfo{year}{2001}).

\bibitem[{\citenamefont{Bangerth et~al.}(2002)\citenamefont{Bangerth, Hartmann,
  and Kanschat}}]{deal}
\bibinfo{author}{\bibfnamefont{W.}~\bibnamefont{Bangerth}},
  \bibinfo{author}{\bibfnamefont{R.}~\bibnamefont{Hartmann}}, \bibnamefont{and}
  \bibinfo{author}{\bibfnamefont{G.}~\bibnamefont{Kanschat}},
  \emph{\bibinfo{title}{{\tt deal.{I}{I}} Differential Equations Analysis
  Library, Technical Reference}}, \bibinfo{organization}{IWR}
  (\bibinfo{year}{2002}),
  \urlprefix\url{http://gaia.iwr.uni-heidelberg.de/~deal/}.

\bibitem[{\citenamefont{Ainsworth and Craig}(1991)}]{ains}
\bibinfo{author}{\bibfnamefont{M.}~\bibnamefont{Ainsworth}} \bibnamefont{and}
  \bibinfo{author}{\bibfnamefont{A.}~\bibnamefont{Craig}},
  \bibinfo{journal}{Numer. Math. 60} pp. \bibinfo{pages}{429--463}
  (\bibinfo{year}{1991}).

\end{thebibliography}

\end{document}